\documentclass[aps,prl,twocolumn,floatfix, preprintnumbers]{revtex4}
\usepackage[pdftex]{graphicx}
\usepackage[mathscr]{eucal}
\usepackage[pdftex]{hyperref}
\usepackage{amsmath,amssymb,bm,yfonts}
\usepackage[utf8]{inputenc}

\newif\ifarxiv
\arxivfalse

\begin		{document}

\def\section	#1{\quad\textit{#1}.---}
\def\G {H}
\def\suck[#1]#2{\includegraphics[#1]{#2}}        

\title
    {
    Numerical evolution of shocks in the interior of Kerr black holes
    }

\author{Paul~M.~Chesler}
\affiliation
    {Black Hole Initiative, Harvard University, Cambridge, MA 02138, USA}
\email{pchesler@g.harvard.edu}

\author{Erik~Curiel}
\affiliation
    {Munich Center for Mathematical Philosophy, Ludwig-Maximilians-Universit\"at,  
    	Ludwigstra\ss \ 31, 80539 M\"unchen, Germany  
    	and
    Black Hole Initiative, Harvard University, Cambridge, MA 02138, USA}
\email{erik@strangebeautiful.com}


\author{Ramesh Narayan}
\affiliation
{Black Hole Initiative, Harvard University, Cambridge, MA 02138, USA }
\email{rnarayan@cfa.harvard.edu}

\date{\today}

\begin{abstract}
We numerically solve Einstein's equations coupled to a scalar field in the interior of Kerr black holes.  
We find shock waves form near the inner horizon. 
The shocks grow exponentially in amplitude and need not be axisymmetric.
Observers who pass through the shocks experience exponentially large tidal forces and are accelerated exponentially close to the speed of light.
\end{abstract}


\maketitle
\parskip	2pt plus 1pt minus 1pt

\section{Introduction}The no-hair theorem postulates 
that the exterior geometry of
black holes is completely described by
the black hole's mass, charge and angular momentum via the 
Kerr-Newman metric.  The origin of this lies in the fact 
that perturbations near the event horizon can either be absorbed by the 
event horizon or radiated to infinity, allowing the near horizon geometry to relax.  In fact
the no-hair theorem should also hold just inside the event horizon as well, since just inside the
event horizon of the Kerr-Newmann metric, all light rays propagate deeper into the interior.  
However, inside the inner horizon of the Kerr-Newmann geometry,
light rays need not propagate deeper into the interior, meaning there is no mechanism for perturbations to relax and the no-hair theorem does not apply there.

The interior geometry of black holes has been most widely studied for Reissner-Nordstr\"om black holes \cite{Penrose:1968ar,Simpson:1973ua,HISCOCK1981110,PhysRevD.20.1260,Poisson:1989zz,PhysRevD.41.1796,PhysRevLett.67.789,0264-9381-10-6-006,Brady:1995ni,Burko:1997zy,Hod:1998gy,Burko:1997fc,Eilon:2016osg,10.2307/3597235,Marolf:2011dj}. This is due to the fact that one can impose spherical symmetry,  greatly simplifying calculations.  
One striking result is the presence of \textit{gravitational shock waves} \cite{Marolf:2011dj,Eilon:2016osg}, which form along the 
outgoing branch of the inner horizon %
\footnote
  {
  More precisely, the shocks form just outside where the outgoing 
  branch of the inner horizon would be in the Reissner-Nordstr\"om
  geometry.
  }.
The formation mechanism of the shocks essentially lies in the fact that all outgoing 
light rays between the event and inner horizons of the Reissner-Nordstr\"om geometry
asymtotote to the inner horizon at late times.  This means that outgoing 
radiation between the inner and event horizons --- which can always be excited via scattering of infalling radiation --- becomes localized to an arbitrarily thin shell as time progresses.  This thin shell of radiation has a dramatic effect on infalling geodesics passing through it.  Firstly, via Raychauhuri's equation, it follows that radial null geodesics propagate from the shock to the central singularity 
over an exponentially small affine parameter.  Likewise, upon passing through the shock, time-like radial observers are dramatically accelerated, 
experiencing exponentially large tidal forces, and encounter the central singularity 
an exponentially short proper time later.  For a solar mass black hole, the proper time interval between passing through the shock and smashing into the central singularity becomes Plackian
milliseconds after the black hole forms. Hence infalling observers encounter a curvature brick wall
at the shock, where the curvature increases from its approximate Reissner-Nordstr\"om value to infinity over
Planckian proper times.

For Kerr black holes (or more generally Kerr-Newman black holes) outgoing light rays between the event and inner horizons also asymptote to the inner horizon at late times.  Hence, it is natural to expect 
gravitational shocks to form in the interior of rotating black holes \cite{Marolf:2011dj}.
In the present work we study the evolution of shocks in Kerr black holes 
by numerically 
solving Einstein's equations coupled to a scalar field. We study both axisymmetric and non-axisymmetric solutions.
Like Reissner-Nordstr\"om black holes, we find shocks form near the outgoing leg of the inner horizon. 
In addition to solving Einstein's equations numerically, we also solve them analytically with a derivative 
expansion in the vicinity of the shocks and find excellent agreement with the numerics.
Like shocks in Reissner-Nordstr\"om black holes, we find that shocks in rotating black holes 
dramatically affect infalling geodesics passing through them.  In particular, infalling time-like 
observers passing through the shocks are  accelerated exponentially close to the speed of light and experience exponentially large tidal forces.

\section{Setup}
We numerically solve Einstein's equations coupled to
a massless real scalar field $\Psi$. The equations of
motion read 
\begin{align}
\label{eq:Einstein}
R^{\mu \nu}- {\textstyle \frac{1}{2} } R g^{\mu \nu} = 8 \pi T^{\mu \nu},
\end{align}
and 
\begin{align}
\label{eq:scalareq}
D^2 \Psi = 0,
\end{align}
where $D_\mu$ is the covariant derivative operator and
\begin{align}
T_{\mu \nu} = D_\mu \Psi D_\nu \Psi - {\textstyle \frac{1}{2}} g_{\mu \nu} (D \Psi)^2,
\end{align}
is the scalar stress tensor.

Our numerical evolution scheme is detailed in 
\cite{Chesler:2013lia}.  Here we outline the salient details.  
We employ a characteristic evolution scheme where the metric takes the form
\begin{equation}
\label{eq:metric}
\! ds^2 = -2 A dv^2 {+} 2 dv d\lambda + r^2 h_{ab} (dx^a {-} F^a dv)(dx^b {-} F^b dv),
\end{equation}
with $x^{a} = \{\theta,\varphi\}$ where $\theta$ is the polar angle and $\varphi$ is the azimuthal angle.  
The two dimensional angular metric $h_{ab}$ satisfies $\det h_{ab} = \sin^2 \theta$, meaning the 
function $r$ is an areal coordinate.  Lines of constant time $v$ and angles $\theta, \varphi$ are radial null infalling geodesics.  The radial coordinate $\lambda$ is an affine parameter for these geodesics.  Correspondingly,
the metric (\ref{eq:metric}) is invariant 
under the residual diffeomorphism
\begin{equation} 
\lambda \to \lambda + \xi(v,\theta,\varphi),
\end{equation}
where $\xi$ is arbitrary.
We fix $\xi$  
such that the inner horizon
of the stationary Kerr geometry is located at 
\begin{equation}
\lambda = \lambda_- = 1.
\end{equation}

\begin{figure}
	\includegraphics[trim= 0 0 0 0 ,clip,scale=0.75]{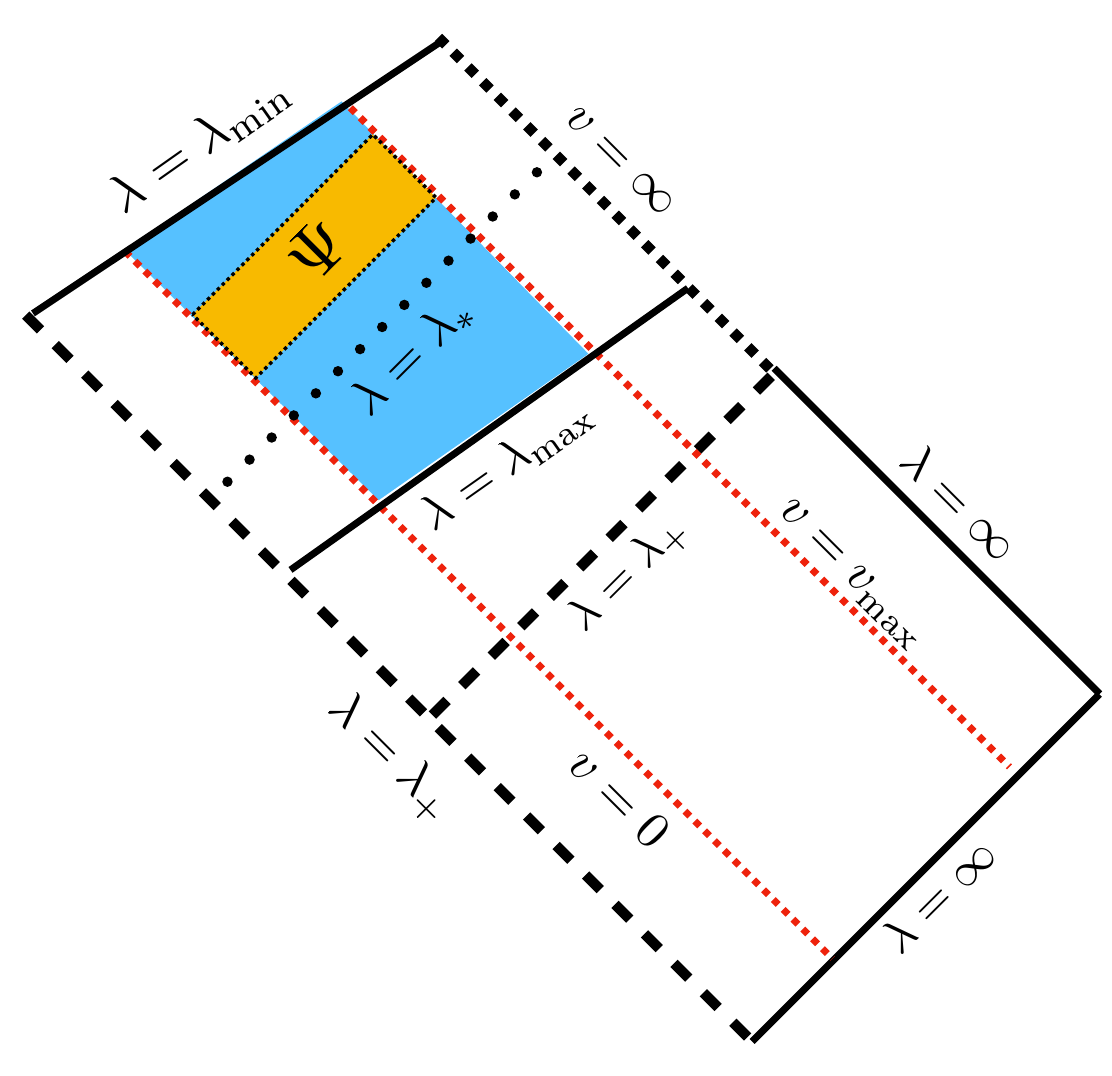}
	\caption{
		A Penrose diagram showing our computational domain, which is represented by the blue shaded region.
	    Our radial coordinate $\lambda$ is the affine parameter for infalling 
		radial geodesics, two of which are shown as the red dashed lines.  The event horizon is located at $\lambda = \lambda_+$.
		The scalar field $\Psi$ (yellow shaded region) is localized inside the event horizon between $\lambda_{\rm min}(v)$ and $\lambda_{\rm max}(v)$,
		which are spacelike surfaces which at late times asymtotote to $\lambda = 1$, the location of the inner horizon of the Kerr geometry.  The outgoing null surface $\lambda = \lambda_*$ is employed as a matching surface in our analytic calculations below.
	}
	\label{fig:Setup}
\end{figure}

\begin{figure}
	\begin{center}
		\includegraphics[trim= 0 0 0 0 ,clip,scale=0.26]{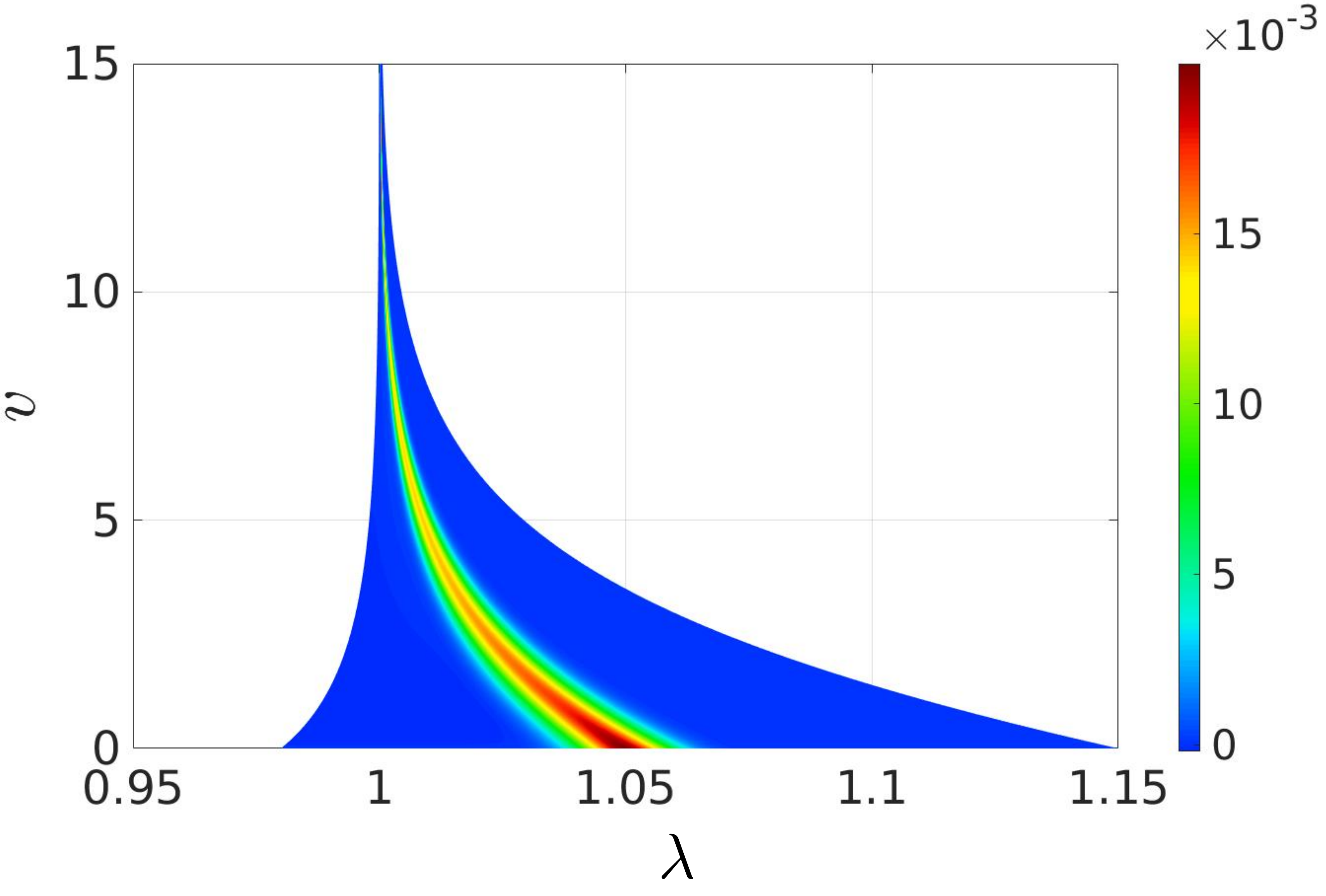}
		\caption{
			Evolution of the scalar field $\Psi$ 
			in the equatorial plane for spin $a = 0.9$.
			The inner and outer boundaries of the shaded region
			represent $\lambda_{\rm min}(v)$ and $\lambda_{\rm max}(v)$.
			As time progresses the scalar field becomes localized at $\lambda = 1$.
		}
		\label{fig:ScalarPlot}
	\end{center}
\end{figure}

Requisite initial data at $v = 0$ consists of the scalar field $\Psi$ and the 
angular metric $h_{ab}$.  The remaining 
components of the metric are determined by 
initial value constraint equations \cite{Chesler:2013lia}.
Perhaps the most natural 
initial data is that where a rotating black hole is formed dynamically via gravitational collapse.  Another  option would be to start with a Kerr initial data and allow infalling radiation to perturb the geometry inside the event horizon at $\lambda = \lambda_+$.  A third
option is to start with Kerr initial data and add a perturbation \textit{inside} the event horizon.
To study the evolution of shocks, 
it is sufficient to consider the last option, as this offers several computational advantages. 
First, limiting perturbations to the interior of the black hole means that one can restrict the computational domain to the interior of the black hole.  Second, since no energy or angular momentum can be radiated to infinity, the mass and spin of the black hole remain constant.  Because the geometry outside the inner horizon should be stable, this means that at late times the position of the inner horizon must approach that of the unperturbed Kerr geometry at $\lambda = 1$. In our coordinate system
this ultimately means that at late 
times one must have $A \to 0$ at $\lambda = 1$.
Having the inner horizon approach constant $\lambda$ is useful, since shocks are expected to form there.

We employ the Kerr metric for initial 
$h_{ab}$. For initial scalar data we choose 
\begin{align}
\label{eq:PsiIC}
\!\!\!\!\!\! \Psi = {\textstyle\frac{1}{50}} e^{-(\lambda-\lambda_0)^2/2 \sigma^2} \left\{1 + \zeta \, {\rm Re} [ y^{10}(\theta,\varphi) {+} y^{11}(\theta,\varphi)] \right\},
\end{align}
where $y^{\ell m}$ are 
spherical harmonics and $\zeta$ is a parameter
 controlling the degree of non-axisymmetry in the initial data.
We choose $\lambda_0$ and $\sigma$ such that $\Psi$ is localized between $\lambda = 1$ and $\lambda = \lambda_+$  and 
exponentially small at our outer computational boundary.
We fix the Kerr mass parameter $M = 1$ and spin $a = 0.9, 0.95$ and $0.99$ and evolve until $v_{\rm max} = 9/\kappa$ with 
$\kappa$ the surface gravity of inner horizon of the unperturbed Kerr black hole.  For a = $0.9$,  $0.95$ we set $(\lambda_0,\sigma) =(1.05,1/150)$ while for $a = 0.99$ we set  $(\lambda_0,\sigma)=(1.01,1/500)$.
For axisymmetric initial data we set $\zeta = 0$ and for non-axisymmetric initial data we set $\zeta = 1/4$.

We employ a time dependent radial computational domain
$\lambda_{\rm min}(v) \leq \lambda \leq \lambda_{\rm max}(v)$. $\lambda_{\rm min}(v)$ and $\lambda_{\rm max}(v)$ will be surfaces which at late times 
asymtotote to $\lambda = 1$ from below and above respectively.
See Fig.~\ref{fig:Setup} for a Penrose diagram illustrating our computational 
domain.  We choose 
\begin{align}
&\!\!\! \frac{d\lambda_{\rm max}}{dv} = 
\min\limits_{\theta, \varphi} 
A \,|_{\lambda = \lambda_{\rm max}},&&
\frac{d\lambda_{\min}}{dv} = {\max\limits_{\theta, \varphi}} \,A \,|_{\lambda = \lambda_{\rm min}}.&
\end{align}
These choices mean that the surfaces $\lambda_{\rm min}(v)$ and $\lambda_{\rm max}(v)$ are either spacelike 
or null.  This in turn means that no information can propagate from inside $\lambda_{\rm min}$ through $\lambda_{\rm min}$.  At 
$\lambda_{\rm max}$, where the scalar field is exponentially small, we impose the boundary condition that the geometry is that of Kerr.  This is allowed since no signal from inside $\lambda_{\rm max}$ can ever reach $\lambda_{\rm max}$. 

Our discretization scheme is nearly identical to that in \cite{Chesler:2018txn}.
To discretize the equations of motion we make a linear change of coordinates from $\lambda$ to $z\in(-1,1)$ via 
\begin{equation}
\lambda = a(v) z + b(v),
\end{equation} 
where 
\begin{subequations}
	\begin{eqnarray}
	a(v) &=& {\textstyle \frac{1}{2}(\lambda_{\rm max}(v) - \lambda_{\rm min}(v))},
	\\
	b(v) &=& {\textstyle \frac{1}{2}(\lambda_{\rm max}(v) + \lambda_{\rm min}(v))}.
	\end{eqnarray}
\end{subequations}
Following \cite{Chesler:2013lia}, we expand the $z$ dependence of all functions in a pseudo-spectral basis 
of Chebyshev polynomials.
We employ domain decomposition 
in $z$ direction with 30 equally spaced domains, each containing 8 points.  

For the $(\theta,\varphi)$ dependence we employ a basis of 
scalar, vector and tensor harmonics.
These are eigenfunctions of the covariant Laplacian $-\nabla^2$ on the unit sphere.
The scalar eigenfunctions are just 
spherical harmonics $y^{\ell m}$.  There are two vector harmonics, $\mathcal V_a^{s \ell m}$ with $s = 1,2$,
and three symmetric tensor harmonics, $\mathcal T_{ab}^{s \ell m}$, $s = 1,2,3$.  Explicit representations 
of these functions are easily found and read \cite{oldref}
\begin{subequations}
	\begin{eqnarray}
	\mathcal V_a^{1\ell m} &=& {\textstyle \frac{1}{\sqrt{\ell (\ell + 1)}}} \nabla_a y^{\ell m}, \\
	\mathcal V_a^{2\ell m} &=&  {\textstyle \frac{1}{\sqrt{\ell (\ell + 1)}}} \epsilon_a^{\ b} \nabla_b y^{\ell m},\\
	\mathcal  T_{ab}^{1 \ell m} &=& {\textstyle \frac{\G_{ab}}{\sqrt{2}}} y^{\ell m}, 
	\\ 
	\mathcal T_{ab}^{2\ell m} &=& {\textstyle \frac{1}{ \sqrt{\ell (\ell + 1)( \ell(\ell + 1)/2 - 1)}}} 
	\epsilon_{(a}^{\ \  c} \nabla_{b)} \nabla_c   y^{\ell m},
	\\
	\mathcal T^{3 \ell m}_{ab} &=& {\textstyle \frac{1}{\sqrt{\ell (\ell + 1)( \ell(\ell + 1)/2 - 1)}}} [ \nabla_a \nabla_b 
	+ {\textstyle\frac{\ell (\ell + 1)}{2}} \G_{ab} ]  y^{\ell m}, \ \ \ \  \ \ \ 
	\end{eqnarray}
\end{subequations}
where $\epsilon_a^{\ b}$ has non-zero components $\epsilon_\theta^{\ \varphi} = \csc \theta$ and $\epsilon_{\varphi}^{\ \theta} = -\sin \theta$, and $\G_{ab} = {\rm diag}(1,\sin^2 \theta)$ is the metric 
on the unit sphere.
The scalar, vector and tensor harmonics are orthonormal and complete.

We expand the metric and scalar field as follows,
\begin{subequations}
	\label{eq:expansions}
	\begin{eqnarray}
	g_{00}(v,z,\theta,\varphi)  &=& \sum_{\ell m}  \alpha^{\ell m}(v,z) y^{\ell m}(\theta,\varphi),
	\\
	g_{0a}(v,z,\theta,\varphi)  &=& \sum_{s\ell m}  \beta^{s\ell m}(v,z) \mathcal V^{s\ell m}_a(\theta,\varphi),
	\\
	g_{ab}(v,z,\theta,\varphi)  &=& \sum_{s\ell m}  \gamma^{s\ell m}(v,z) \mathcal  T^{s\ell m}_{ab}(\theta,\varphi), 
	\\
	\Psi(v,z,\theta,\varphi) &=&\sum_{\ell m}  \chi^{\ell m}(v,z) y^{\ell m}(\theta,\varphi).
	\end{eqnarray}
\end{subequations}
Derivatives in $\{\theta,\varphi\}$ can be taken by differentiating the scalar, vector and tensor harmonics.

In order to efficiently transform between real space and mode space, we employ a Gauss-Legendre grid in $\theta$ with $\ell_{\rm max} + 1$ points.  Likewise, we employ a Fourier grid in the $\varphi$ direction with $2 \ell_{\rm max} + 1$ points.
These choices allow the transformation between mode space and real space to be done with a combination 
of Gaussian quadrature and Fast Fourier Transforms.

We truncate the expansions (\ref{eq:expansions}) at maximum angular momentum $\ell_{\rm max} = 100$.  
For axisymmetric simulations we also truncate at azimuthal
quantum number $m_{\rm max} = 0$.  For non-axisymmetric simulations we truncate at $m_{\rm max} = 20$.

 \begin{figure}
	\includegraphics[trim= 0 0 0 0,clip,scale=0.25]{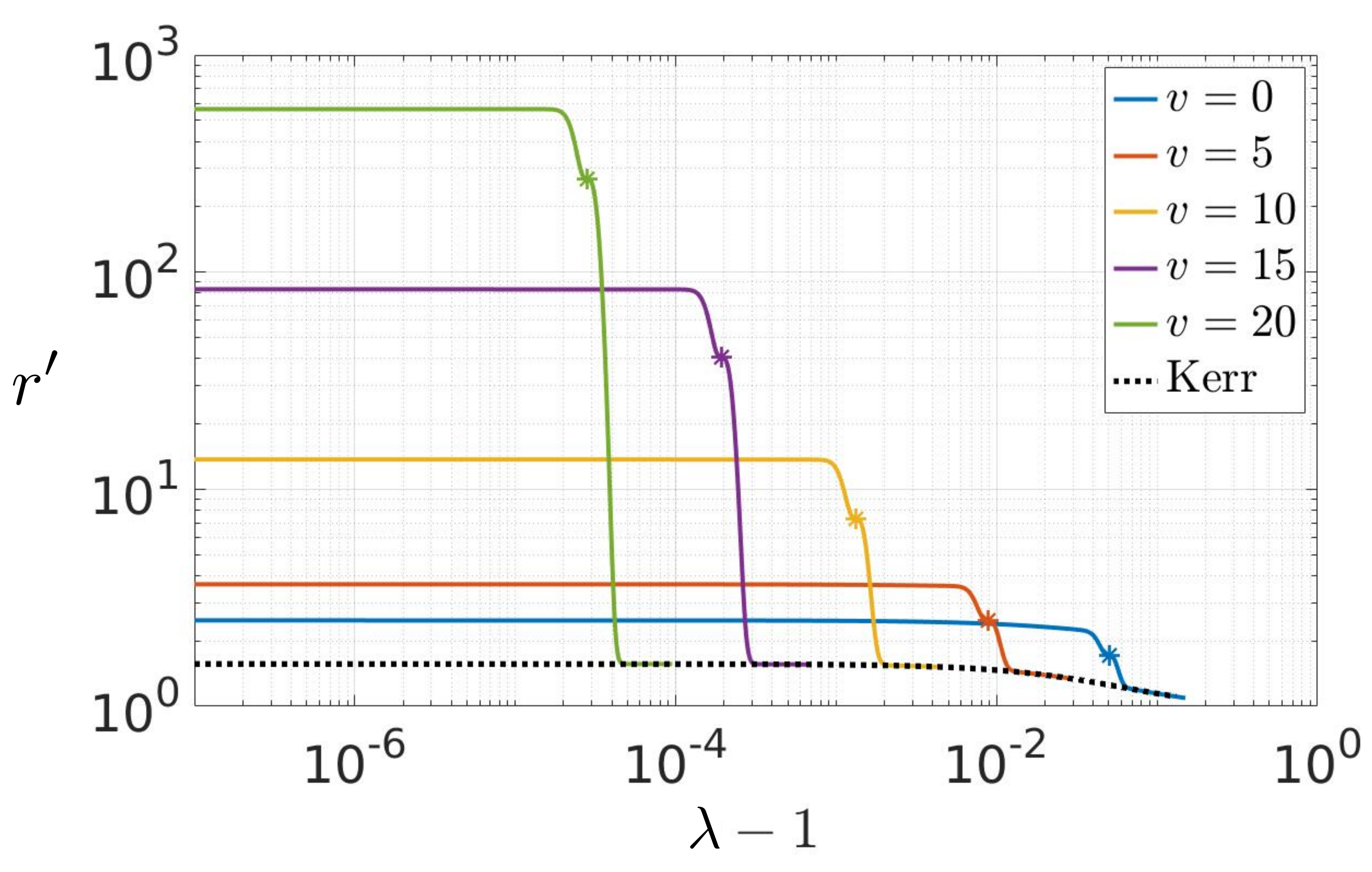}
	\caption{
		$r'$ in the equatorial plane at several times for $a = 0.9$. The $*$ denote location of the maximum of $\Psi$ 
		at the corresponding time.  A shock in $r'$ is evident.  Outside the shock, $r'$ approaches its Kerr value while inside $r'$ grows in time. 
	}
	\label{fig:SigmaPrime}
\end{figure}

\begin{figure*}
	
	\includegraphics[trim= 10 0 0 10 ,clip,scale=0.55]{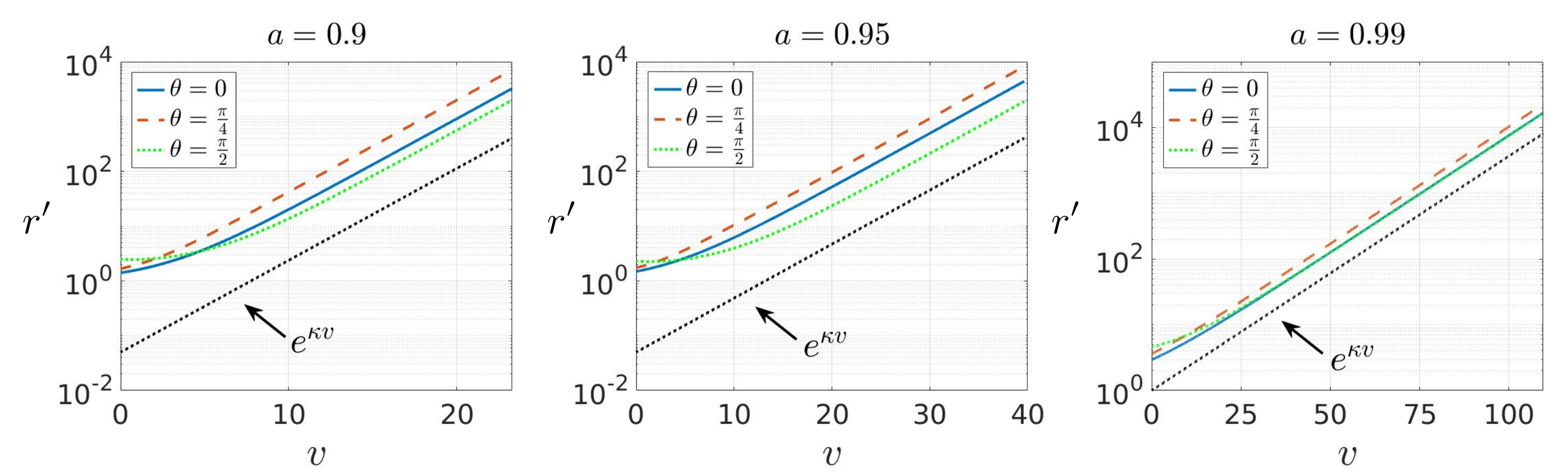}
	\caption{
    $r'|_{\lambda = 1}$ for axisymmetric simulations with 
    spin $a = 0.9, 0.95$ and $0.99$.
	}
	\label{fig:SigmaPrimeScaling}
\end{figure*}

\section{Results and discussion}
We begin by presenting results for 
axisymmetric simulations.
In Fig.~\ref{fig:ScalarPlot} we plot the scalar field
$\Psi$ as a function of time $v$ and radial coordinate
$\lambda$ in the equatorial plane for spin $a = 0.9$.  
The inner and outer 
boundaries of the shaded region correspond to the curves $\lambda_{\rm min}(v)$ and $\lambda_{\rm max}(v)$ and reflect our time-dependent computational domain. 
As time progresses the scalar wave packet propagates inwards towards $\lambda = 1$, becoming increasingly narrower in the process while
staying roughly constant in magnitude.
As the scalar wave packet approaches $\lambda = 1$, the 
metric at $\lambda > 1$ approaches that of Kerr.

The localization of the scalar wave packet to $\lambda = 1$
results in large $\lambda$ derivatives of the metric at $\lambda \leq 1$.  A useful
metric component to study is the areal coordinate $r$, which 
is related to the volume element via 
$\sqrt{-g} = r^2 \sin \theta$.  In Fig.~\ref{fig:SigmaPrime} we plot $r'|_{\lambda = 1}$ in the equatorial plane at several times, again for spin $a = 0.9$. Here and below 
\begin{equation}
' \equiv \partial_\lambda.
\end{equation}
The $*$ in the figure denote the maximum of $\Psi$ at the corresponding time. As is evident from the figure, 
there is a dramatic change in $r'$ near the 
scalar maxima.  In other words, there is a shock in $r'$.  Exterior to the shock $r'$ is well approximated by its Kerr value.  The change in $r'$ across the shock grows with time. 

In Fig.~\ref{fig:SigmaPrimeScaling} we plot
$r'|_{\lambda = 1}$ as a function of $v$ for several values of $\theta$
and for $a = 0.9, 0.95$ and $0.99$.  Also
included in each plot is $e^{\kappa v}$ where
\begin{equation}
\kappa = \frac{1}{2} \left ( \frac{1}{M - \sqrt{M^2 - a^2}}  - \frac{1}{M}\right),
\end{equation}
is the surface gravity
of the inner horizon of the corresponding Kerr solution.
For $a = 0.9, 0.95$ and $0.99$ we have $\kappa \approx 0.386, 0.227$ and $0.0821$, respectively.  Our numerics are consistent with the scaling $r' \sim e^{\kappa v}$.

\begin{figure}[h!]
	\begin{center}
		\includegraphics[trim= 0 0 0 0 ,clip,scale=0.25]{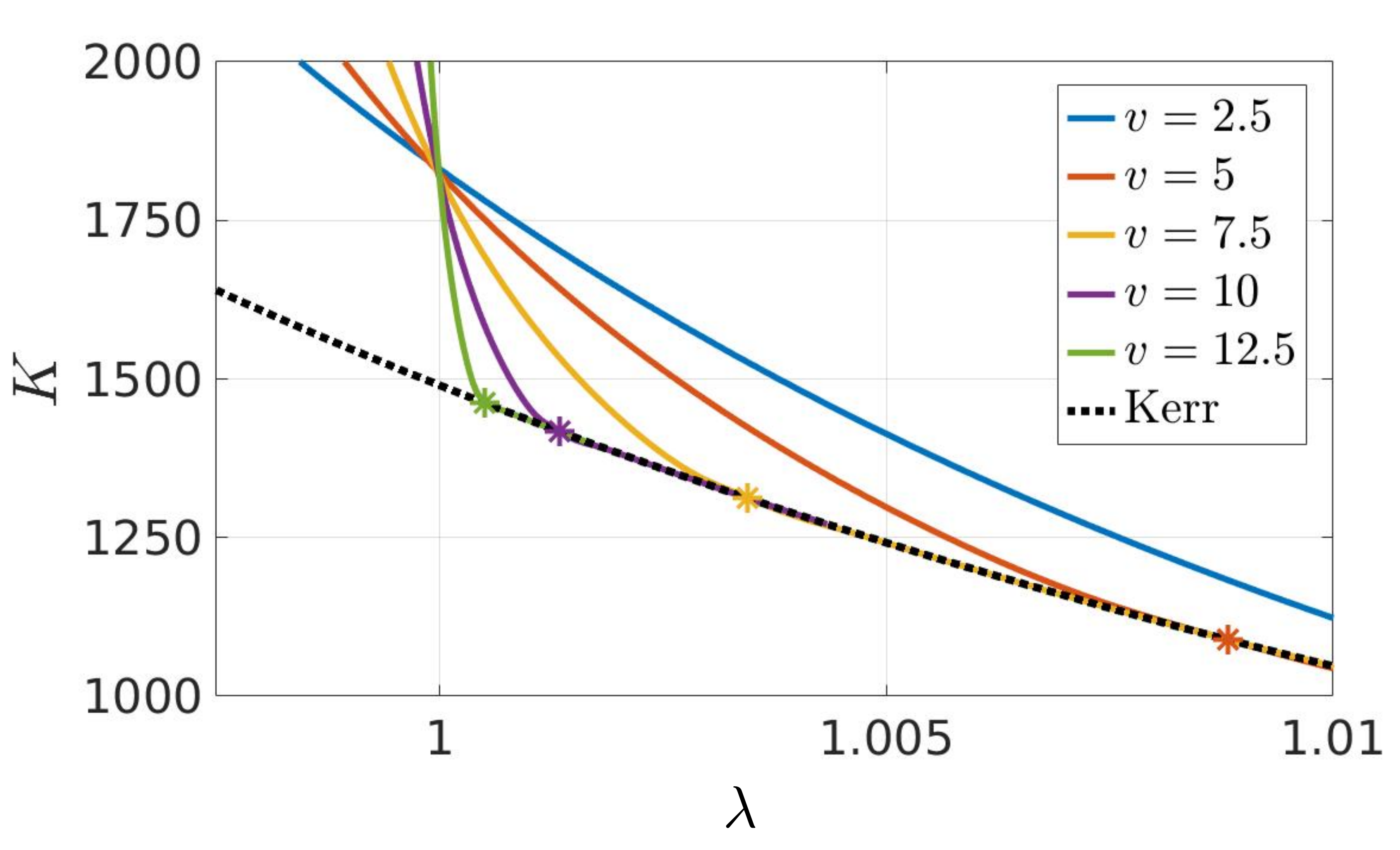}
		\caption{The Kretschmann scalar $K$
			 in the equatorial plane at several times for an axisymmetric simulation with $a = 0.9$. The $*$ denote location of the maximum of the scalar wave packet  
			at the corresponding time. Exterior to the wave packet $K$ is well described by its Kerr value.  At $\lambda = 1$ $K$ is nearly constant but $K'$ grows with time.
		}
		\label{fig:curvscalar}
	\end{center}
\end{figure}

\begin{figure*}[ht]
	\begin{center}
		\includegraphics[trim= 180 10 100 10 ,clip,scale=0.19]{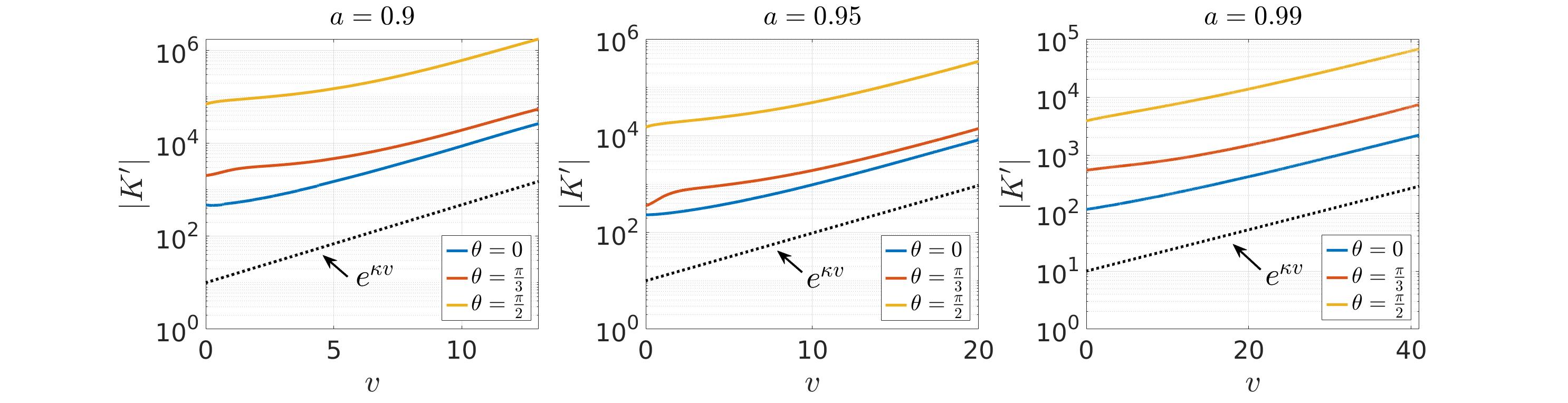}
		\caption{
		$|K'||_{\lambda = 1}$ evaluated 
		at several polar angles for axisymmetric simulations with 
		spin $a = 0.9, 0.95$ and $0.99$.	
		}
		\label{fig:curvderscaling}
	\end{center}
\end{figure*}

We now turn to the curvature.  
In Fig.~\ref{fig:curvscalar} we plot the Kretschmann  scalar 
\begin{equation}
\label{eq:K}
K \equiv R^{\mu \nu \alpha \beta} R_{\mu \nu \alpha \beta},
\end{equation}
as a function of $\lambda$ in the equatorial plane at several times for the same simulation shown in Fig.~\ref{fig:SigmaPrimeScaling}.
The $*$ denote the location of the maximum of $\Psi$ at the corresponding time. Exterior to the scalar wave packet, $K$ is well approximated by its Kerr value.  
A prominent feature of Fig.~\ref{fig:curvscalar} is that $K'$ grows dramatically with time just inside the wave packet.
In Fig.~\ref{fig:curvderscaling} we plot $|K'| |_{\lambda = 1}$ 
as a function of $v$ at several values of $\theta$ for the same simulations shown in 
Fig.~\ref{fig:SigmaPrimeScaling}. Also included in the plots is $e^{\kappa v}$.  Our numerics are consistent with the scaling $|K'| |_{\lambda = 1} \sim e^{\kappa v}$.
Evidently, $\lambda = 1$ becomes a \textit{curvature brick wall}
at late times, with a shock in $K$ developing there. 

The geometry 
in the vicinity of the shocks can be studied perturbatively \cite{Marolf:2011dj}. To 
account for the rapid $\lambda$ dependence, we introduce a bookkeeping 
parameter $\epsilon \ll 1$ and assume the scalings
\begin{align}
\label{eq:scalings}
&\partial_\lambda \sim 1/\epsilon,&
A \sim \epsilon.&
\end{align}
The scaling $A \sim \epsilon$ is necessary to have $A' = O(1)$,
which itself is necessary to have a finite surface gravity at $\lambda = 1$ as $v \to \infty$.
We then solve the Einstein-scalar system in the region $\lambda < \lambda_*$ with $\lambda_*   = 1+O( \epsilon)$.  A convenient choice of matching surface $\lambda = \lambda_*$ is an outgoing null sheet exterior to the 
scalar wavepacket, as shown in Fig.~\ref{fig:Setup}.

On the surface $\lambda = \lambda_*$
we impose the boundary condition that the geometry is that of Kerr and that the scalar field vanishes.
At leading order in $\epsilon$ we therefore need the inner horizon values of the Kerr metric.
In our coordinate system, at the inner horizon of Kerr we have 
\begin{align}
\label{eq:bc1}
&A|_{\lambda =1} = 0, & & A'|_{\lambda =1} = -\kappa,&& F^a|_{\lambda =1} = \Omega \delta^{a \phi},&
\end{align}
where 
\begin{equation}
\Omega= \frac{\sqrt{M^2 - a^2}+M}{2 a M},
\end{equation}
is the angular 
velocity of the inner horizon.
It follows that for the Kerr solution 
\begin{equation}
\label{eq:bc2}
D_+ g_{\mu \nu}|_{\lambda =1} = 0,
\end{equation}
where 
\begin{equation}
\label{eq:Dplus}
D_+ \equiv \partial_v + A \partial_\lambda + F^a \partial_a,
\end{equation}
is the directional derivative along outgoing null geodesics.

At leading order in $\epsilon$ the $(\mu,\nu) = (v,a)$ component of the Einstein equations 
(\ref{eq:Einstein}) reduces to
\begin{align}
\label{eq:Feq}
F'^a = 0.
\end{align}
The boundary conditions
(\ref{eq:bc1}) therefore implies
\begin{equation}
\label{eq:Fsol}
F^a = \Omega \delta^{a \phi}.
\end{equation}
With the solution (\ref{eq:Fsol}), at leading order in $\epsilon$ the 
$(\mu,\nu) = (v,\lambda)$ and $(a,b)$ components of Einstein's equations reduce to %
\footnote
 {
 The remaining 
 components of Einstein's equations, the $(\mu, \nu) = (v,v), \ (\lambda,\lambda)$ and $(\lambda,a)$ components,
 are initial value and radial constraint equations, respectively, and will not be necessary for our analysis here.	
 }
\begin{subequations}
\begin{align}
\label{eq:rdot}
0& = (r D_+ r)',
\\
\label{eq:hdot}
0&= \textstyle \left [ \delta^{c}_{a} \delta^{d}_b - \frac{1}{2}h^{cd} h_{ab}\right ] (r D_+ h_{cd})'+D_+ r h'_{ab},
\\ \label{eq:Aeq}
0&= \textstyle A'' + \frac{1}{4} h^{ab} h'_{bc} h^{cd} D_+ h_{d a} - \frac{2 r'D_+ r}{r}  
\\ \nonumber
&+ 8 \pi \Psi' D_+ \Psi, 
\end{align}
\end{subequations}
where $h^{ab}$ is the inverse of $h_{ab}$. Likewise, at leading order in $\epsilon$ the scalar equation of motion (\ref{eq:scalareq}) reduces to
\begin{align}
\label{eq:scalarwave}
(r D_+ \Psi) = - \Psi' D_+ r.
\end{align}

Eqs.~(\ref{eq:rdot}), (\ref{eq:hdot}) and (\ref{eq:scalarwave}) are just 
radial wave equations for $r$, $h_{ab}$ and $\Psi$.  Imposing the boundary 
conditions that the geometry exterior to the shell is that of Kerr is tantamount to 
imposing the boundary conditions that there is no infalling radiation through the shell, which
is what (\ref{eq:bc2}) states.  With the boundary condition (\ref{eq:bc2}), 
Eqs.~(\ref{eq:rdot}), (\ref{eq:hdot}) and (\ref{eq:scalarwave}) have the solutions
\begin{align}
\label{eq:firstordersysm}
&D_+ r = 0, & & D_+ \Psi = 0, && D_+ h_{ab} = 0.&
\end{align}
These first order wave equations state that excitations in $r$, $\Psi$ and $h_{ab}$ are transported 
along outgoing null geodesics tangent to $D_+$.  Substituting (\ref{eq:firstordersysm}) into (\ref{eq:Aeq}) and employing the boundary conditions (\ref{eq:bc1}), we secure 
\begin{equation}
\label{eq:Asol}
A = -\kappa \lambda.
\end{equation}

With the solutions (\ref{eq:Asol}) and (\ref{eq:Fsol}) and the definition of $D_+$ in (\ref{eq:Dplus}),
the first order system (\ref{eq:firstordersysm}) is solved by 
\begin{subequations}
	\label{eq:geooptics}
	\begin{align}
	r(v,\lambda,\theta,\varphi) &= \rho(e^{\kappa v}(\lambda - 1),\theta,\varphi -\Omega v), \\
	\Psi(v,\lambda,\theta,\varphi) &= \psi(e^{\kappa v}(\lambda - 1),\theta,\varphi -\Omega v), \\
	h_{ab}(v,\lambda,\theta,\varphi) &= \mathcal H_{ab}(e^{\kappa v}(\lambda - 1),\theta,\varphi -\Omega v),
	\end{align}
\end{subequations}
where $\rho$, $\psi$ and $\mathcal H_{ab}$ are arbitrary functions. 
Note that curves with $e^{\kappa v}(\lambda - 1)$, $\theta,$ and $\varphi -\Omega v$ all constant are simply outgoing 
null geodesics near $\lambda = 1$.  
These geodesics 
spiral in towards $\lambda = 1$ at angular frequency $\Omega$, which is due to frame dragging, and eventually terminate 
at $\lambda = 1$ as $v \to \infty$.  The value of the fields on these geodesics is constant.  Since the $\lambda$ dependence comes in the combination $e^{\kappa v} (\lambda - 1)$, it follows that $e^{-\kappa v}$ plays the role 
of our bookkeeping parameter $\epsilon$.

\begin{figure}
	\includegraphics[trim= 
	0 0 0 40,clip,scale=0.2]{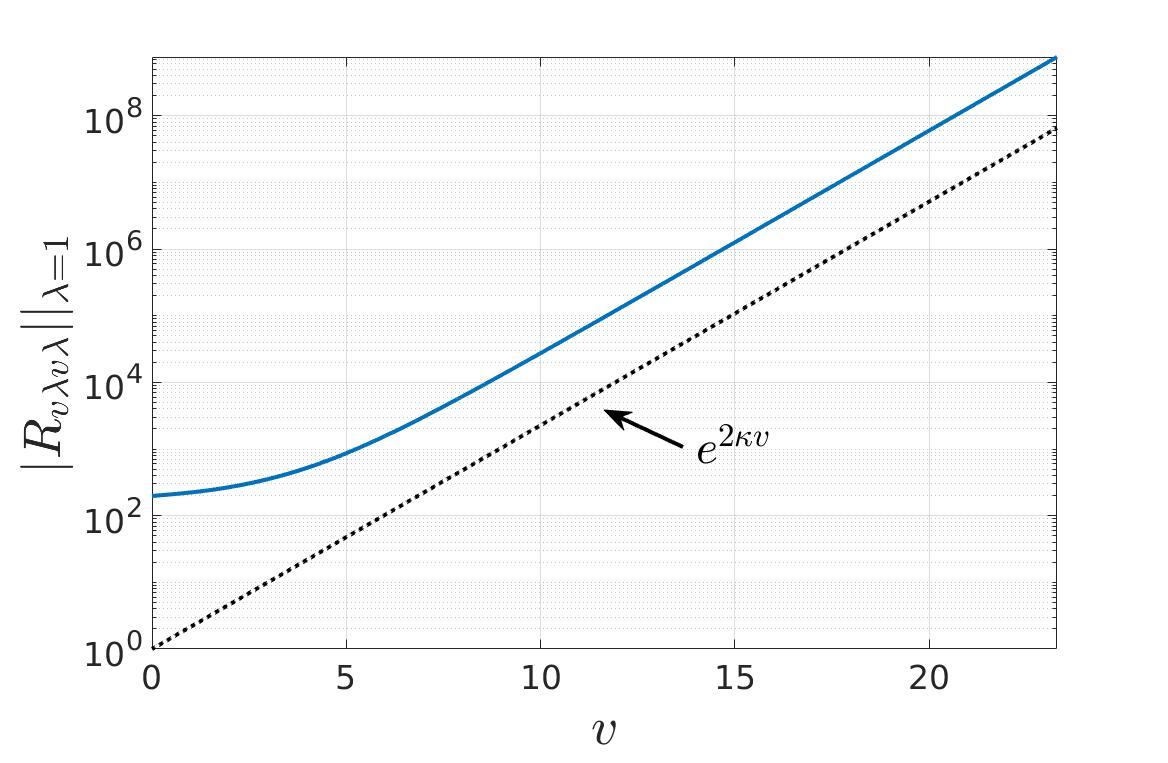}
	\caption{The Riemann tensor component $R_{v \lambda v \lambda}|_{\lambda = 1}$ in the equatorial plane for an axisymmetric simulations with spin $a = 0.9.$  The Riemann tensor diverges like $e^{2 \kappa v}$.
	}
	\label{fig:RiemannScalings}
\end{figure}

The above analysis implies  
that as 
$v \to \infty$, the scalar wave packet must approach $\lambda = 1$, just as seen in Fig.~\ref{fig:ScalarPlot}.
Moreover, it immediately follows from (\ref{eq:geooptics}) that the individual components of the 
Riemann tensor scale like 
\begin{equation}
\label{eq:Riemann}
R_{\mu \nu \alpha \beta} \sim e^{2 \kappa v}.
\end{equation}
Indeed, in Fig.~\ref{fig:RiemannScalings} we plot the component $R_{v \lambda v \lambda}|_{\lambda =1}$
in the equatorial plane for an axisymmetric simulation with spin $a = 0.9$ and 
verify this scaling.
The exponential growth in (\ref{eq:Riemann}) reflects the fact that outgoing radiation 
is blue shifted by a factor of $e^{\kappa v}$, becoming 
exponentially localized in $\lambda$ in the process.  However, owing to the fact that
all excitations in (\ref{eq:geooptics}) are purely outgoing, the curvature scalar $K$
cannot blow up exponentially, meaning all exponential factors in (\ref{eq:K}) cancel.  Why must this happen?  Since by construction there is no infalling radiation present, one
can simply boost to the frame where the outgoing radiation is not blue shifted and the components $R_{\mu \nu \alpha \beta}$ and Kretschmann scalar are finite as $v \to \infty$. Simply put, with only outgoing radiation present, the Kretschmann scalar --- and in fact all other scalars --- can only depend on $\{v,\lambda,\varphi\}$ through the combinations $e^{\kappa v}(\lambda - 1)$ and $\varphi - \Omega v$.  It therefore follows that $K$ is finite on the shocks and that 
\begin{align}
\label{eq:3Dscalings}
&r'|_{\lambda = 1} =e ^{\kappa v} H(\theta,\varphi - \Omega v), &
K'|_{\lambda = 1} = e^{\kappa v} Q(\theta,\varphi - \Omega v),
\end{align}
for some functions $H$ and $Q$. The scaling relations (\ref{eq:3Dscalings}) match those shown in Figs.~\ref{fig:SigmaPrimeScaling} and \ref{fig:curvscalar} for 
our axisymmetric simulations.

\begin{figure}
	\includegraphics[trim= 60
	0 0 0 40,clip,scale=0.17]{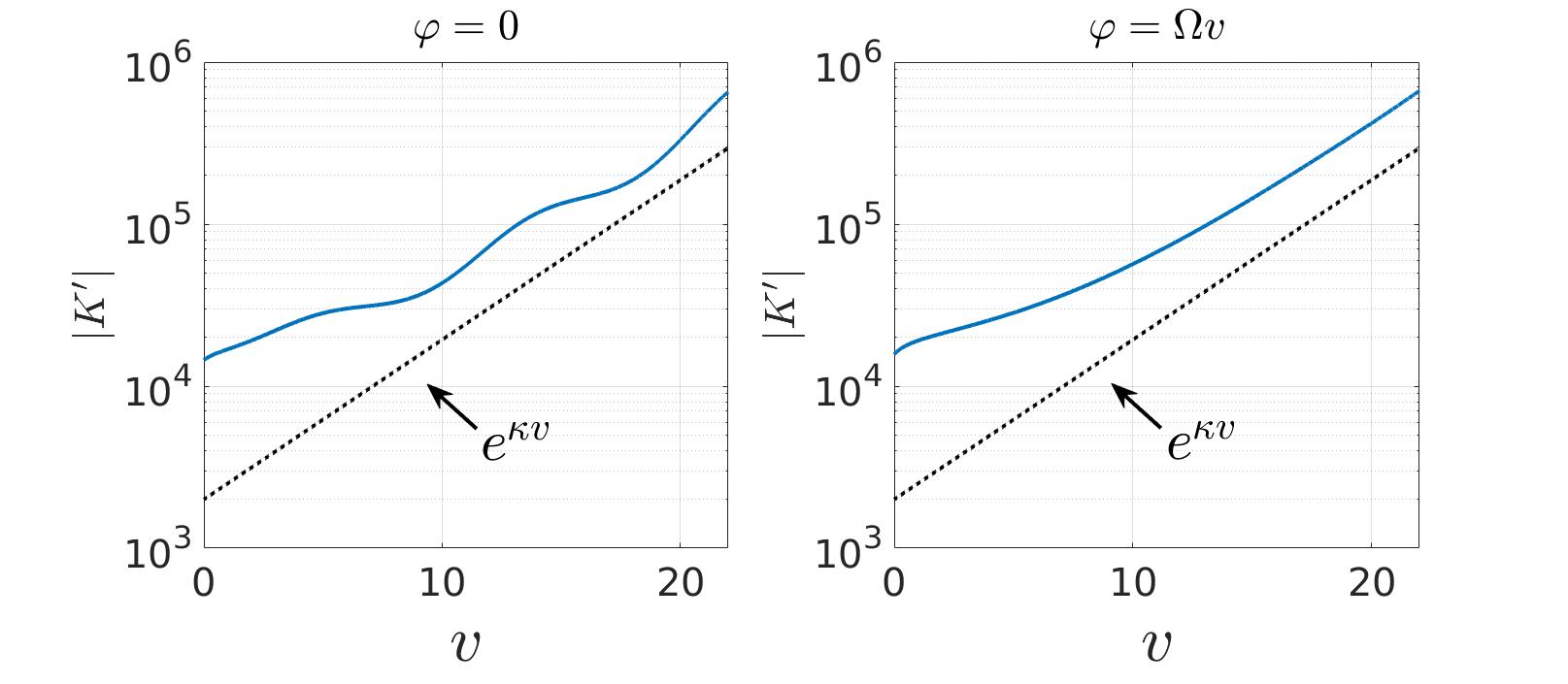}
	\caption{$|K'||_{\lambda = 1}$ in the equatorial plane for a non-axisymmtric simulation with spin $a = 0.95$.  The left
		plot is evaluated at $\varphi = 0$ whereas the 
		right is at $\varphi = \Omega v$.  
	}
	\label{fig:K3DScalings}
\end{figure}

The scaling relations
(\ref{eq:3Dscalings}) also
demonstrate rotation invariance in $\varphi$ can be broken: a small non-axisymmetric perturbation in initial data results in violations of axisymmetry in $r'$ and $K'$ which are exponentially amplified.
To demonstrate this, in Fig.~\ref{fig:K3DScalings}
we plot $K'|_{\lambda = 1}$ at $\theta = \pi/2$ as a function of time  
for a non-axisymmetric simulation with $a = 0.95$.
The left figure is evaluated at $\varphi =0$ while the right figure is evaluated at $\varphi = \Omega v$.  At $\varphi = 0$ we 
see that $K'$ grows exponentially with 
sinusoidal oscillations superimposed.
In the rotating frame, where $\varphi = \Omega v$, the sinusoidal oscillations are not present, just as (\ref{eq:3Dscalings}) requires. 
Evidently, the curvature brick wall at $\lambda = 1$ 
retains angular structure contained in the initial data.
Oscillating features of the curvature were also reported in \cite{Ori:2001pc}.

Let us now turn to analyzing the effect of the shocks 
on infalling geodesics.  Consider first radial infalling null geodesics with $v = {\rm const}.$  
The null energy condition
implies that $r''/r < 0$.  This means that $r'$ can only \textit{increase} as $\lambda$ decreases.
Since just inside the shock $r' \sim e^{\kappa v}$, it follows that the 
affine distance $\Delta \lambda$ from the shock to the point $r = 0$ is 
\begin{equation}
\label{eq:Dlambda}
\Delta \lambda \sim e^{-\kappa v}.
\end{equation}
This reflects the fact that the shock \textit{focuses}
infalling light rays to $r = 0$, which can also 
be seen from Raychaudhuri's equation.
Turning now to infalling time-like geodesics, the scaling (\ref{eq:Riemann}) means that infalling time-like observers crossing the shocks will experience tidal forces of order $e^{2 \kappa v}$.  In particular, upon
crossing the shocks the areal velocity will be 
\begin{equation}
\frac{dr}{d\tau} \sim -e^{\kappa v}.
\end{equation}
What then is the fate of an observer who jumps into the black hole at late times?  For large enough black holes, observers need not experience any ill effects until they pass through the shocks.  They will
measure the local geometry to be that of Kerr, with arbitrarily small tidal forces.  Upon encountering the shocks though, they will be torn apart by tidal forces and their subsequent debris will be accelerated nearly to the speed of light towards the black hole interior.

In the present paper we only considered perturbations in the interior of black holes and did not allow infalling radiation.  Exterior perturbations of black holes in asymtotically flat spacetime results in infalling radiation which decays with a power law in $v$ in accords with Price's Law \cite{Price:1972pw}.
For Reissner-Nordstr\"om black holes, 
infalling radiation results in a weak null curvature singularity developing on the ingoing leg of the inner horizon ($v = \infty$ in Fig.~\ref{fig:Setup}) \cite{Poisson:1989zz,PhysRevD.41.1796}.  A similar effect should happen for Kerr black holes.  With infalling radiation, the exponential factors in (\ref{eq:Riemann}) cannot be ameliorated via a boost, for a boost which compensates the $e^{\kappa v}$ blueshift of outgoing radiation will inevitably result
in the infalling radiation being blue shifted by a factor of $e^{\kappa v}$.  With infalling radiation present, it is therefore reasonable to expect that $K$ will blow up like $e^{2 \kappa v}$.  We leave the inclusion of infalling radiation for future studies.

\section{Acknowledgments}%
This work was supported by the Black Hole Initiative at Harvard University, 
which is funded by a grant from the John Templeton Foundation. EC is also supported by grant 312032894 from the
Deutsche Forschungsgemeinschaft.  We thank Amos Ori and Peter Galison
for useful comments and discussions.

\bibliographystyle{utphys}
\bibliography{refs}%

\providecommand{\href}[2]{#2}\begingroup\raggedright\begin{thebibliography}{10}

\bibitem{Penrose:1968ar}
R.~Penrose,
``{Structure of space-time},''.

\bibitem{Simpson:1973ua}
M.~Simpson and R.~Penrose, ``{Internal instability in a Reissner-Nordstrom
  black hole},''
\href{http://dx.doi.org/10.1007/BF00792069}{{\em Int. J. Theor. Phys.} {\bf 7}
  (1973)  183--197}.

\bibitem{HISCOCK1981110}
W.~A. Hiscock, ``Evolution of the interior of a charged black hole,''
  \href{http://dx.doi.org/https://doi.org/10.1016/0375-9601(81)90508-9}{{\em
  Physics Letters A} {\bf 83} (1981) no.~3, 110 -- 112}.
  \url{http://www.sciencedirect.com/science/article/pii/0375960181905089}.

\bibitem{PhysRevD.20.1260}
Y.~G\"ursel, I.~D. Novikov, V.~D. Sandberg, and A.~A. Starobinsky,
  \href{http://dx.doi.org/10.1103/PhysRevD.20.1260}{``Final state of the
  evolution of the interior of a charged black hole,''{\em Phys. Rev. D} {\bf
  20} (Sep, 1979)  1260--1270}.
  \url{https://link.aps.org/doi/10.1103/PhysRevD.20.1260}.

\bibitem{Poisson:1989zz}
E.~Poisson and W.~Israel, ``{Inner-horizon instability and mass inflation in
  black holes},''
\href{http://dx.doi.org/10.1103/PhysRevLett.63.1663}{{\em Phys. Rev. Lett.}
  {\bf 63} (1989)  1663--1666}.

\bibitem{PhysRevD.41.1796}
E.~Poisson and W.~Israel,
  \href{http://dx.doi.org/10.1103/PhysRevD.41.1796}{``Internal structure of
  black holes,''{\em Phys. Rev. D} {\bf 41} (Mar, 1990)  1796--1809}.
  \url{https://link.aps.org/doi/10.1103/PhysRevD.41.1796}.

\bibitem{PhysRevLett.67.789}
A.~Ori, \href{http://dx.doi.org/10.1103/PhysRevLett.67.789}{``Inner structure
  of a charged black hole: An exact mass-inflation solution,''{\em Phys. Rev.
  Lett.} {\bf 67} (Aug, 1991)  789--792}.
  \url{https://link.aps.org/doi/10.1103/PhysRevLett.67.789}.

\bibitem{0264-9381-10-6-006}
M.~L. Gnedin and N.~Y. Gnedin, ``Destruction of the cauchy horizon in the
  reissner-nordstrom black hole,'' {\em Classical and Quantum Gravity} {\bf 10}
  (1993) no.~6, 1083. \url{http://stacks.iop.org/0264-9381/10/i=6/a=006}.

\bibitem{Brady:1995ni}
P.~R. Brady and J.~D. Smith, ``{Black hole singularities: A Numerical
  approach},'' \href{http://dx.doi.org/10.1103/PhysRevLett.75.1256}{{\em Phys.
  Rev. Lett.} {\bf 75} (1995)  1256--1259},
\href{http://arxiv.org/abs/gr-qc/9506067}{{\tt arXiv:gr-qc/9506067 [gr-qc]}}.

\bibitem{Burko:1997zy}
L.~M. Burko, ``{Structure of the black hole's Cauchy horizon singularity},''
  \href{http://dx.doi.org/10.1103/PhysRevLett.79.4958}{{\em Phys. Rev. Lett.}
  {\bf 79} (1997)  4958--4961},
\href{http://arxiv.org/abs/gr-qc/9710112}{{\tt arXiv:gr-qc/9710112 [gr-qc]}}.

\bibitem{Hod:1998gy}
S.~Hod and T.~Piran, ``{Mass inflation in dynamical gravitational collapse of a
  charged scalar field},''
  \href{http://dx.doi.org/10.1103/PhysRevLett.81.1554}{{\em Phys. Rev. Lett.}
  {\bf 81} (1998)  1554--1557},
\href{http://arxiv.org/abs/gr-qc/9803004}{{\tt arXiv:gr-qc/9803004 [gr-qc]}}.

\bibitem{Burko:1997fc}
L.~M. Burko and A.~Ori, ``{Analytic study of the null singularity inside
  spherical charged black holes},''
  \href{http://dx.doi.org/10.1103/PhysRevD.57.R7084}{{\em Phys. Rev.} {\bf D57}
  (1998)  7084--7088},
\href{http://arxiv.org/abs/gr-qc/9711032}{{\tt arXiv:gr-qc/9711032 [gr-qc]}}.

\bibitem{Eilon:2016osg}
E.~Eilon and A.~Ori, ``{Numerical study of the gravitational shock wave inside
  a spherical charged black hole},''
  \href{http://dx.doi.org/10.1103/PhysRevD.94.104060}{{\em Phys. Rev.} {\bf
  D94} (2016) no.~10, 104060},
\href{http://arxiv.org/abs/1610.04355}{{\tt arXiv:1610.04355 [gr-qc]}}.

\bibitem{10.2307/3597235}
M.~Dafermos, ``Stability and instability of the cauchy horizon for the
  spherically symmetric einstein-maxwell-scalar field equations,'' {\em Annals
  of Mathematics} {\bf 158} (2003) no.~3, 875--928.
  \url{http://www.jstor.org/stable/3597235}.

\bibitem{Marolf:2011dj}
D.~Marolf and A.~Ori, ``{Outgoing gravitational shock-wave at the inner
  horizon: The late-time limit of black hole interiors},''
  \href{http://dx.doi.org/10.1103/PhysRevD.86.124026}{{\em Phys. Rev.} {\bf
  D86} (2012)  124026},
\href{http://arxiv.org/abs/1109.5139}{{\tt arXiv:1109.5139 [gr-qc]}}.

\bibitem{Chesler:2013lia}
P.~M. Chesler and L.~G. Yaffe, ``{Numerical solution of gravitational dynamics
  in asymptotically anti-de Sitter spacetimes},''
  \href{http://dx.doi.org/10.1007/JHEP07(2014)086}{{\em JHEP} {\bf 07} (2014)
  086},
\href{http://arxiv.org/abs/1309.1439}{{\tt arXiv:1309.1439 [hep-th]}}.

\bibitem{Chesler:2018txn}
P.~M. Chesler and D.~A. Lowe, ``{Nonlinear evolution of the AdS$_4$ black hole
  bomb},''
\href{http://arxiv.org/abs/1801.09711}{{\tt arXiv:1801.09711 [gr-qc]}}.

\bibitem{oldref}
V.~D. Sandberg, ``Tensor spherical harmonics on $s^2$ and $s^3$ as eigenvalue
  problems,'' \href{http://dx.doi.org/10.1063/1.523649}{{\em Journal of
  Mathematical Physics} {\bf 19} (1978) no.~12, 2441--2446},
  \href{http://arxiv.org/abs/https://doi.org/10.1063/1.523649}{{\tt
  https://doi.org/10.1063/1.523649}}.

\bibitem{Ori:2001pc}
A.~Ori, ``{Oscillatory null singularity inside realistic spinning black
  holes},'' \href{http://dx.doi.org/10.1103/PhysRevLett.83.5423}{{\em Phys.
  Rev. Lett.} {\bf 83} (1999)  5423--5426},
\href{http://arxiv.org/abs/gr-qc/0103012}{{\tt arXiv:gr-qc/0103012 [gr-qc]}}.

\bibitem{Price:1972pw}
R.~H. Price, ``{Nonspherical Perturbations of Relativistic Gravitational
  Collapse. II. Integer-Spin, Zero-Rest-Mass Fields},''
\href{http://dx.doi.org/10.1103/PhysRevD.5.2439}{{\em Phys. Rev.} {\bf D5}
  (1972)  2439--2454}.

\end{thebibliography}\endgroup

\end{document}